\documentclass[preprint]{aastex}

\newcommand       \be           {\begin{equation}}
\newcommand       \ee           {\end{equation}}
\newcommand       \Angstrom     {\,{\rm \AA}}
\newcommand       \eV           {\,{\rm eV}\,}

\newcommand       \cm           {\,{\rm cm}}
\newcommand       \s            {\,{\rm s}}

\newcommand       \erg          {\,{\rm erg}}

\newcommand       \ltsim        {\lesssim}

\shortauthors{Weingartner \& Jordan}
\shorttitle{Torques on Grains}

\begin{document}

\title{Torques on Spheroidal Silicate Grains Exposed to Anisotropic 
Interstellar Radiation Fields}

\author{Joseph C. Weingartner \& Margaret E. Jordan}
\affil{Department of Physics and Astronomy, George Mason University,
MSN 3F3, 4400 University Drive, Fairfax, VA 22030, USA}

\begin{abstract}

Radiative torques, due to the absorption and scattering of starlight, are
thought to play a major role in the alignment of grains with the interstellar
magnetic field.  The absorption of radiation also gives rise to recoil 
torques, associated with the photoelectric effect and photodesorption.
The recoil torques are much more difficult to model and compute than the 
direct radiative torque.  Here, we consider the relatively simple case
of a spheroidal grain.  Given our best estimates for the photoelectric yield
and other relevant grain physical properties, we find that the recoil torques 
contribute at the $\approx 10\%$ level or less compared with the direct 
radiative torque.  We recommend that the recoil torques not be included in 
models of radiation-driven grain alignment at this time.  However, additional 
experimental characterization of the surface properties and photoelectric 
yield for sub-micron grains is needed to better quantify the magnitude of
these torques.

\end{abstract}

\keywords{ISM: dust}

\section{\label{sec:intro} Introduction}

As starlight traverses the dusty interstellar medium (ISM), it acquires a
partial linear polarization, due to dichroic extinction.  Grains are
non-spherical and are more effective at attenuating radiation polarized
along their ``long'' axes, which tend to align along the same direction
in space.  Despite decades of investigation, the alignment mechanism is
still not fully understood; see Lazarian (2003), Roberge (2004), and
Lazarian (2007) for recent reviews.  

Radiative torques, due to the absorption and scattering of starlight, 
appear to play a major role in grain alignment 
(Draine \& Weingartner 1996, 1997).  If the incident radiation field is 
anisotropic, then radiative torques can directly align grains, at least
for the few grain shapes that have been investigated so far.  
A detailed investigation, incorporating a large suite of shapes, is needed
in order to fully assess the viabilty of radiation-driven alignment.
Although the alignment is not due to magnetic torques in this scenario, 
the observed polarization is nevertheless 
correlated with the interstellar magnetic field direction, since the 
grains have magnetic dipole moments lying along their spin axes and thus 
precess rapidly about the field direction (Martin 1971;  
Dolginov \& Mytrophanov 1976).

In addition to the direct torque due to absorption and scattering,
there are also recoil torques, associated with photoelectric emission and 
photodesorption. In the latter process, adsorbed atoms or molecules
(i.e., gas-phase particles that have stuck to the grain surface after 
colliding with it) are ejected back into the gas after an absorbed photon 
breaks the bond between the adsorbed particle and the surface.
So far, photoelectric and photodesorption torques associated with
anisotropic radiation have not been considered.
However, one expects that they could be significant compared with the 
direct radiative torque, since Weingartner \& Draine (2001a) found that
the photoelectric and photodesorption {\it forces} can be comparable to 
the direct radiation pressure force in some interstellar environments.

Here, we estimate the photoelectric torque for two simple grain shapes, 
namely, prolate and oblate spheroids with axis ratios of 1.5.  We also
estimate the maximum magnitude for the photodesorption torque.
We consider grains with silicate composition and $a_{\rm eff} = 0.2 \micron$
($a_{\rm eff}$ is the radius of a sphere with equal volume); such grains are
known to be well-aligned in the diffuse ISM (e.g., Kim \& Martin 1995).

Photoelectrons typically originate within $\sim 10 \Angstrom$ of the grain
surface and the ejection rate is proportional to the electric intensity
$|\mathbf{E}|^2$.  
Thus, the field inside the grain must be determined to within high resolution
($10 \Angstrom$ out of $0.2 \micron$) in order to compute the photoelectric
torque, unless the near-surface field remains fairly constant on larger
length scales.  
Draine \& Weingartner (1996, 1997) computed radiative torques using the
discrete dipole approximation, in which the grain is approximated as a set
of polarizable points.  The size of the dipole array needed to achieve a
resolution of $10 \Angstrom$ out of $0.2 \micron$ is prohibitive.  
Thus, we adopt the point matching method (PMM), which can achieve high 
resolution at relatively low computational cost.
In this approach, the internal and scattered fields are expanded in vector
spherical wave functions and the expansion coefficients are found by imposing 
boundary conditions at a set of discrete points on the grain surface.

The PMM simplifies considerably for grains with azimuthal symmetry; this 
motivated our choice of spheroids.  However, these grains are too symmetric
for radiation torques (including the recoil torques) to yield alignment, if
the surface properties are uniform.  
The purpose of this paper is simply to estimate the magnitude of the recoil
torques relative to the direct radiative torque, in order
to determine whether or not recoil 
torques need to be included in future studies of radiation-driven alignment.
Detailed scattering computations are required even for this simple goal, 
since the wavelength of the incident radiation is comparable to the grain 
size.  In this case, the internal electric intensity cannot be simply 
estimated using geometrical optics and an attenuation coefficient.  

In \S \ref{sec:pmm}, we describe our implementation of the PMM, which is 
based on the treatments of Morrison \& Cross (1974) and Al-Rizzo \& 
Tranquilla (1995).   
\S \ref{sec:q} describes the computation of efficiency factors for 
extinction, scattering, radiation pressure, and radiative torque.  
Readers who are interested only in the results should skip \S \S 
\ref{sec:pmm} and \ref{sec:Q_formalism}.
In \S \ref{sec:pe}, we describe our simple model for the photoelectric force
and torque and present computational results.  \S \ref{sec:photodesorption}
briefly examines the photodesorption torque.  We conclude 
(\S \ref{sec:conclusion}) with a discussion of the implications of these
estimates for future grain alignment studies.

\section{\label{sec:pmm} Point Matching Method}

In spherical coordinates, solutions of the vector Helmholtz equation are
given by $\mathbf{M}_{nm}(r, \theta, \phi)$ and 
$\mathbf{N}_{nm}(r, \theta, \phi)$, with
\be
M_r(r, \theta, \phi) = 0
\ee
\be
M_{\theta}(r, \theta, \phi) = \frac{im}{\sin \theta} \, Z(kr) \, 
P(\cos \theta) \, e^{im\phi}
\ee
\be
M_{\phi}(r, \theta, \phi) = \sin \theta \, Z(kr) \, P^{\prime}(\cos \theta) 
\, e^{im\phi}
\ee
\be
N_r(r, \theta, \phi) = \frac{n \, (n+1)}{kr} \, Z(kr) \, P(\cos \theta) 
\, e^{im\phi}
\ee
\be
N_{\theta}(r, \theta, \phi) = - \sin \theta \, \left[ Z^{\prime}(kr) + 
\frac{Z(kr)}{kr} \right] \, P^{\prime}(\cos \theta) \, e^{im \phi}
\ee
\be
N_{\phi}(r, \theta, \phi) = \frac{im}{\sin \theta} \, \left[ Z^{\prime}(kr) 
+ \frac{Z(kr)}{kr} \right] \, P(\cos \theta) \, e^{im \phi}
\ee
where $P(\cos \theta) = P_n^{|m|}(\cos \theta)$, 
$P^{\prime}(\cos \theta) = dP_n^{|m|}(\cos \theta)/d(\cos \theta)$,
$k$ is the wave number ($k = k_0 \equiv 2 \pi / \lambda$ outside the 
grain and $k = n_{\rm ref} k_0$ inside the grain, where $n_{\rm ref}$ is
the grain index of refraction and $\lambda$ is the wavelength of the incident
radiation), 
$Z(kr) = j_n(k r)$ (spherical Bessel function of the first kind) inside the 
grain and $Z(kr) = h_n^{(1)}(k_0 r)$ (spherical Hankel function of the first 
kind) outside the grain, and $Z^{\prime}(kr) = dZ(kr)/d(kr)$.  We adopt
Jackson's (1999) convention for associated Legendre polynomials
$P_n^m(\cos \theta)$.

The scattered electric ($\mathbf{E}_{\rm sca}$) and magnetic
($\mathbf{H}_{\rm sca}$) fields are given by
\be
\label{eq:esca}
\mathbf{E}_{\rm sca} = \sum_{n=1}^{\infty} \sum_{m=-n}^n 
\left[ a_{nm}(1) \, \mathbf{N}_{nm}^{(s)} + a_{nm}(2) \, \mathbf{M}_{nm}^{(s)} 
\right] 
\ee
\be
\label{eq:hsca}
\mathbf{H}_{\rm sca} = - i \sum_{n=1}^{\infty} \sum_{m=-n}^n 
\left[ a_{nm}(1) \, \mathbf{M}_{nm}^{(s)} + a_{nm}(2)
\, \mathbf{N}_{nm}^{(s)} \right]
\ee
and the internal fields are given by
\be
\label{eq:eint}
\mathbf{E}_{\rm int} = \sum_{n=1}^{\infty} \sum_{m=-n}^n 
\left[ a_{nm}(3) \, \mathbf{N}_{nm}^{(w)} + a_{nm}(4)
\, \mathbf{M}_{nm}^{(w)} \right]
\ee
\be
\label{eq:hint}
\mathbf{H}_{\rm int} = - i n_{\rm ref} \sum_{n=1}^{\infty} \sum_{m=-n}^n
\left[ a_{nm}(3) \, \mathbf{M}_{nm}^{(w)} + a_{nm}(4) \, \mathbf{N}_{nm}^{(w)} 
\right]
\ee
where superscripts $(s)$ and $(w)$ refer to evaluation in the grain exterior
and interior, respectively.

We will take the $z$-axis as the symmetry axis of the spheroidal grain and
assume that the incident radiation field wave vector $\mathbf{k}_0$ lies in 
the $x$-$z$ plane with 
$\mathbf{k}_0 = k_0 (\cos \theta_0 \, \hat{z} - \sin \theta_0
\, \hat{x})$.  We also adopt unit amplitude for the incident electric field
$\mathbf{E}_{\rm inc}$.  When $\mathbf{E}_{\rm inc}$ lies 
in the $x$-$z$ plane (TM mode), 
\be
\mathbf{E}_{\rm inc, \, {\rm TM}} = [ \hat{r} (\cos \theta_0 \sin \theta \cos 
\phi + \sin \theta_0 \cos \theta) + \hat{\theta} (\cos \theta_0 \cos \theta 
\cos \phi - \sin \theta_0 \sin \theta) - \hat{\phi} \cos \theta_0 \sin \phi] 
\, \exp(i \mathbf{k}_0 \cdot \mathbf{r})~~~,
\ee
\be
\mathbf{H}_{\rm inc, \, {\rm TM}} = [ \hat{r} \sin \theta \sin \phi + 
\hat{\theta} \cos \theta \sin \phi + \hat{\phi} \cos \phi] 
\, \exp(i \mathbf{k}_0 \cdot \mathbf{r})
\ee 
with 
\be
\mathbf{k}_0 \cdot \mathbf{r} = k_0 r (\cos \theta_0 \cos \theta - \sin 
\theta_0 \sin \theta \cos \phi)~~~.
\ee
For the orthogonal polarization state (TE mode), 
\be
\mathbf{E}_{\rm inc, \, {\rm TE}} = \mathbf{H}_{\rm inc, \, {\rm TM}} 
\ \ \ ; \ \ \ 
\mathbf{H}_{\rm inc, \, {\rm TE}} = - \mathbf{E}_{\rm inc, \, {\rm TM}}~~~.
\ee

At the surface, the tangential components of $\mathbf{E}$ and $\mathbf{H}$ and
the normal components of $\mathbf{D} \equiv \epsilon \mathbf{E}$ ($\epsilon$
is the dielectric function) and $\mathbf{H}$ are continuous.  

For the grain surface, the radius $r$ as a function of polar angle $\theta$
is given by
\be
r(\theta) = a \left[ \cos^2 \theta + \zeta^2 \sin^2 \theta \right]^{-1/2}~~~.
\ee
For prolate grains, $a$ is the semimajor axis length and $\zeta > 1$ is the 
ratio of the semimajor axis length to the semiminor axis length. 
For oblate grains, $a$ is the semiminor axis length and
$\zeta < 1$ is the ratio of the semiminor axis length to semimajor axis 
length.  Given the radius $a_{\rm eff}$ of a sphere
of equal volume, $a = a_{\rm eff} \, \zeta^{2/3}$.

In this case, the boundary conditions on the tangential field components
become
\be
\label{eq:bndrye1}
E_{{\rm inc}, \phi} + E_{{\rm sca}, \phi} = E_{{\rm int}, \phi}
\ee
\be
\label{eq:bndryh1}
H_{{\rm inc}, \phi} + H_{{\rm sca}, \phi} = H_{{\rm int}, \phi}
\ee
\be
\label{eq:bndrye2}
(1 - \zeta^2) \sin \theta \cos \theta \, (E_{{\rm inc}, r} + 
E_{{\rm sca}, r} - E_{{\rm int}, r}) + (\cos^2 \theta + \zeta^2
\sin^2 \theta) \, (E_{{\rm inc}, \theta} + E_{{\rm sca}, \theta} - 
E_{{\rm int}, \theta}) = 0
\ee
\be
\label{eq:bndryh2}
(1 - \zeta^2) \sin \theta \cos \theta \, (H_{{\rm inc}, r} +
H_{{\rm sca}, r} - H_{{\rm int}, r}) + (\cos^2 \theta + \zeta^2
\sin^2 \theta) \, (H_{{\rm inc}, \theta} + H_{{\rm sca}, \theta} -
H_{{\rm int}, \theta}) = 0~~~.
\ee
The boundary conditions on the normal field components are
\be
\label{eq:bndrye3}
(\cos^2 \theta + \zeta^2 \sin^2 \theta) \, (D_{{\rm inc}, r} + 
 D_{{\rm sca}, r} -  D_{{\rm int}, r}) + (\zeta^2 -1) \sin \theta \cos \theta 
\, (D_{{\rm inc}, \theta} + D_{{\rm sca}, \theta} - D_{{\rm int}, \theta}) = 0
\ee
\be
\label{eq:bndryh3}
(\cos^2 \theta + \zeta^2 \sin^2 \theta) \, (H_{{\rm inc}, r} + 
 H_{{\rm sca}, r} -  H_{{\rm int}, r}) + (\zeta^2 -1) \sin \theta \cos \theta 
\, (H_{{\rm inc}, \theta} + H_{{\rm sca}, \theta} - H_{{\rm int}, \theta}) = 
0~~~.
\ee

The incident radiation field can be expressed as follows:
\be
\label{eq:eincr}
E_{r, {\rm TM}} = \exp(i k_0 r \cos \theta_0 \cos \theta) \, 
\sum_{m= - \infty}^{\infty} i^m \, [ J_m(u) \sin \theta_0 \cos
\theta - i J_m^{\prime}(u) \cos \theta_0 \sin \theta ] \, \exp(i m \phi)
\ee
\be
\label{eq:einctheta}
E_{\theta, {\rm TM}} = - \exp(i k_0 r \cos \theta_0 \cos \theta) \,
\sum_{m= - \infty}^{\infty} i^m \, [ J_m(u) \sin \theta_0 \sin \theta + i 
J_m^{\prime}(u) \cos \theta_0 \cos \theta] \, \exp(i m \phi)
\ee
\be
\label{eq:eincphi}
E_{\phi, {\rm TM}} = \exp(i k_0 r \cos \theta_0 \cos \theta) \, \cos \theta_0 
\, \sum_{m= - \infty}^{\infty} i^m m u^{-1} J_m(u) \exp(i m \phi)
\ee
\be
\label{eq:hincr}
H_{r, {\rm TM}} = - \sin \theta \, \exp(i k_0 r \cos \theta_0 \cos \theta) \,
\sum_{m= - \infty}^{\infty} i^m m u^{-1} J_m(u) \exp(i m \phi)
\ee
\be
\label{eq:hinctheta}
H_{\theta, {\rm TM}} = - \cos \theta \, \exp(i k_0 r \cos \theta_0 \cos 
\theta) \, \sum_{m= - \infty}^{\infty} i^m m u^{-1} J_m(u) \exp(i m \phi)
\ee
\be
\label{eq:hincphi}
H_{\phi, {\rm TM}} = - \exp(i k_0 r \cos \theta_0 \cos \theta) \,
\sum_{m= - \infty}^{\infty} i^{m+1} J_m^{\prime}(u) \exp(i m \phi)
\ee
with $u = - k_0 r \sin \theta_0 \sin \theta$, $J_m(u)$ the Bessel function,
and $J_m^{\prime}(u) = dJ_m(u)/du$.

To find the expansion coefficients $a_{nm}$,
the field expressions in equations (\ref{eq:esca}) through (\ref{eq:hint}) 
and (\ref{eq:eincr}) through (\ref{eq:hincphi}) are substituted into the 
boundary conditions (\ref{eq:bndrye1}) through (\ref{eq:bndryh2}).
In the resulting equations, each term contains a factor $\exp(i m \phi)$
and no other $\phi$-dependence.  Multiplying by $\exp(-i m^{\prime} \phi)$
and integrating over $\phi$, we find the following equations, which much 
be satisfied for each value of $m$, $\theta$, and $i = 1$ through 4:
\be
\label{eq:matrix}
\sum_{n = |m|}^{\infty} \sum_{j=1}^4 C_{nm}(i, j) \cdot a_{nm}(j) = D_m(i)
\ee
with 
\be
C_{nm}(1,1) = (1 - \zeta^2) \sin \theta \cos \theta N_{nm, r}^{(s)}(\phi = 0) 
+ (\cos^2 \theta +  \zeta^2 \sin^2 \theta) N_{nm, \theta}^{(s)}(\phi = 0)
\ee
\be
C_{nm}(1,2) = (\cos^2 \theta + \zeta^2 \sin^2 \theta) M_{nm, \theta}^{(s)}
(\phi = 0)
\ee
\be
C_{nm}(1,3) = - (1 - \zeta^2) \sin \theta \cos \theta 
N_{nm, r}^{(w)}(\phi = 0) - (\cos^2 \theta +
\zeta^2 \sin^2 \theta) N_{nm, \theta}^{(w)}(\phi = 0)
\ee
\be
C_{nm}(1,4) = - (\cos^2 \theta + \zeta^2 \sin^2 \theta) M_{nm, \theta}^{(w)}
(\phi = 0)
\ee
\be
C_{nm}(2,1) = N_{nm, \phi}^{(s)}(\phi = 0)
\ee
\be
C_{nm}(2,2) = M_{nm, \phi}^{(s)}(\phi = 0)
\ee
\be
C_{nm}(2,3) = - N_{nm, \phi}^{(w)}(\phi = 0)
\ee
\be
C_{nm}(2,4) = - M_{nm, \phi}^{(w)}(\phi = 0)
\ee
\be
C_{nm}(3,1) = C_{nm}(1,2)
\ee
\be
C_{nm}(3,2) = C_{nm}(1,1)
\ee
\be
C_{nm}(3,3) = n_{\rm ref} C_{nm}(1,4)
\ee
\be
C_{nm}(3,4) = n_{\rm ref} C_{nm}(1,3)
\ee
\be
C_{nm}(4,1) = C_{nm}(2,2)
\ee
\be
C_{nm}(4,2) = C_{nm}(2,1)
\ee
\be
C_{nm}(4,3) = n_{\rm ref} C_{nm}(2,4)
\ee
\be
C_{nm}(4,4) = n_{\rm ref} C_{nm}(2,3)
\ee
and
\be
\label{eq:d1}
D_m(1) = i^m [J_m(u) \sin \theta_0 \sin \theta \zeta^2 + i J_m^{\prime}(u) 
\cos \theta_0 \cos \theta] \exp(i k_0 r \cos \theta_0 \cos \theta)
\ee
\be
D_m(2) = - i^m m u^{-1} \cos \theta_0 J_m(u) 
\exp(i k_0 r \cos \theta_0 \cos \theta) 
\ee
\be
D_m(3) = i^{m+1} m u^{-1} J_m(u) \cos \theta
\exp(i k_0 r \cos \theta_0 \cos \theta)
\ee
\be
\label{eq:d4}
D_m(4) = - i^m J_m^{\prime}(u) \exp(i k_0 r \cos \theta_0 \cos \theta)~~~.
\ee

In practice, we impose an upper cutoff for $|m|$, $m_{\rm max}$, and for $n$,
$n_{\rm max}$.  For each $m$, we find the expansion coefficients $a_{nm}$
by performing a least-squares minimization of 
\be
\left\| \sum_{\theta, i} \left\{ \sum_{n = |m|}^{n_{\rm max}} \sum_j
\left[ C_{nm}(i, j; \theta) \cdot
a_{nm}(j) - D_m(i; \theta) \right] \right\} \right\|^2
\ee
We generally take $2 n_{\rm max}$ to $3 n_{\rm max}$ values of $\theta$, 
evenly spaced in $\cos \theta$.  

The design matrix for the least-squares problem is a $4 n_{\theta}$ by
$4 (n_{\rm max} - |m| + 1)$ matrix, made up from $C_{nm}(i, j; \theta)$, 
and does not depend on $\theta_0$.  
We first apply a QR-decomposition to the design matrix.  The computational
time required to then complete the solution is small compared with the time
to accomplish the factorization.  Thus, several additional values of 
$\theta_0$ can be treated without substantially increasing the computational
time.  
The matrix equations often exhibit serious ill-conditioning.
We work around this problem by implementing the solution using 
Mathematica, which supports arbitrary-precision arithmetic.  

Note that, when $m$ changes sign, $C(1,2)$, $C(1,4)$, $C(2,1)$, $C(2,3)$,
$C(3,1)$, $C(3,3)$, $C(4,2)$, $C(4,4)$, $D(2)$, and $D(3)$ change sign, 
while the other components retain their sign.  Thus, for the TM mode,
\be
a_{n, -m}(1) = a_{n, m}(1) \ \ ; \ \ 
a_{n, -m}(2) = - a_{n, m}(2) \ \ ; \ \ 
a_{n, -m}(3) = a_{n, m}(3) \ \ ; \ \ 
a_{n, -m}(4) = - a_{n, m}(4) \ \ \ \ {\rm (TM)}
\ee
and for the TE mode, 
\be
a_{n, -m}(1) = - a_{n, m}(1) \ \ ; \ \
a_{n, -m}(2) = a_{n, m}(2) \ \ ; \ \
a_{n, -m}(3) = - a_{n, m}(3) \ \ ; \ \
a_{n, -m}(4) = a_{n, m}(4) \ \ \ \ {\rm (TE)}
\ee
As a result, 
it is not necessary to separately compute the expansion coefficients
for $m < 0$.  When $m=0$, $a_{n, m}(2)$ and $a_{n, m}(4)$ equal zero for
the TM mode and $a_{n, m}(1)$ and $a_{n, m}(3)$ equal zero for the TE mode.

When $\theta_0 = 0$, all of the azimuthal modes vanish, except for those
with $m = \pm 1$.  In this case, 
the above expressions for the incident field components
(eqs.~\ref{eq:eincr}--\ref{eq:hincphi}) contain the ill-defined terms
$J_m^{\prime}(u)$ and $J_m(u)/u$ with $u=0$.  Equations (\ref{eq:d1}) 
through (\ref{eq:d4}) must be replaced with
\be
D_1(1) = D_1(3) = - \frac{1}{2} \cos \theta \exp (i k_0 r \cos \theta)
\ee
\be
D_1(2) = D_1(4) = - \frac{i}{2} \exp (i k_0 r \cos \theta)~~~.
\ee

To verify the solution, we check that the boundary conditions 
(\ref{eq:bndrye1}) through (\ref{eq:bndryh3}) are indeed satisfied.  
This is particularly powerful for conditions (\ref{eq:bndrye3}) and 
(\ref{eq:bndryh3}), since these are not used in the solution.  In 
addition, we check that the field intensity exhibits the following 
symmetry:  $|\mathbf{E}|^2(2 \pi - \phi) = |\mathbf{E}|^2(\phi)$, 
and that the intensity is independent of $\phi$ when $\theta_0 = 0$.

Figures \ref{fig:oblate.0.1.int.4} through \ref{fig:oblate.1.0.ext.4}
show the internal and external (incident plus scattered) fields at the 
grain surface, for a couple illustrative cases.

\section{Efficiency Factors \label{sec:q}}

In this section, we describe the computation of efficiency factors for
scattering, extinction, force, and torque.  The calculation of the direct
radiative torque is needed for comparison with the photoelectric and 
photodesorption torques.  The other efficiency factors will be used to 
validate our PMM code (\S \ref{sec:Q_results}).

\subsection{Definitions and Formulae \label{sec:Q_formalism}}

The scattering efficiency factor $Q_{\rm sca}$ is defined by
\be
\label{eq:Qsca}
P_{\rm sca} = F_{\rm inc} Q_{\rm sca} \pi a_{\rm eff}^2
\ee
where $P_{\rm sca}$ is the scattered power and $F_{\rm inc}$ is the 
incident flux.  The scattered power is found by integrating the flux in 
the scattered fields over a sphere at infinity, yielding
(Morrison \& Cross 1974; Al-Rizzo \& Tranquilla 1995)
\be
Q_{\rm sca} = \frac{4}{(k_0 a_{\rm eff})^2} \sum_{n, m}^{m \ge 0} 
\epsilon_m \, \frac{n (n+1)}{2n+1} \frac{(n+m)!}{(n-m)!} \left[ 
|a_{nm}(1)|^2 + |a_{nm}(2)|^2 \right]~~~;
\ee
$\epsilon_m = 1$ (2) for $m=0$ ($m > 0$).  

The extinction efficiency factor $Q_{\rm ext}$ is defined in the same way
as $Q_{\rm sca}$ (eq.~\ref{eq:Qsca}), except that $P_{\rm sca}$ is replaced
by $P_{\rm abs} + P_{\rm sca}$, where $P_{\rm abs}$ is the absorbed power.
From the optical theorem, 
\be
Q_{\rm ext} = \frac{4}{(k_0 a_{\rm eff})^2} {\rm Im} 
\left\{ \sum_{n, m}^{m \ge 0}
\epsilon_m (-i)^n (-1)^m \left[ a_{nm}(1) \sin \theta_0 P^{\prime}(\cos 
\theta_0) - \frac{m a_{nm}(2) P(\cos \theta_0)}{\sin \theta_0} \right] 
\right\}~~~({\rm TM})
\ee
\be
Q_{\rm ext} = \frac{4}{(k_0 a_{\rm eff})^2} {\rm Im} 
\left\{ \sum_{n, m}^{m \ge 0}
i \epsilon_m (-i)^n (-1)^m \left[ a_{nm}(2) \sin \theta_0 P^{\prime}(\cos
\theta_0) - \frac{m a_{nm}(1) P(\cos \theta_0)}{\sin \theta_0} \right]
\right\}~~~({\rm TE})
\ee
(Al-Rizzo \& Tranquilla 1995).  The absorption efficiency factor
$Q_{\rm abs} = Q_{\rm ext} - Q_{\rm sca}$.  

The force, or pressure, efficiency factor $\mathbf{Q}_{\rm pr}$ is defined by
\be
\label{eq:Q_pr}
\mathbf{F}_{\rm rad} = \pi a_{\rm eff}^2 u_{\rm rad} \mathbf{Q}_{\rm pr} 
\ee
where $\mathbf{F}_{\rm rad}$ is the force on the grain and $u_{\rm rad}$ is 
the incident radiation field energy density.
The torque efficiency factor $\mathbf{Q}_{\Gamma}$ is defined by
\be
\label{eq:Q_Gamma}
\mathbf{\Gamma}_{\rm rad} = \pi a_{\rm eff}^2 u_{\rm rad} \frac{\lambda}{2\pi}
\mathbf{Q}_{\Gamma} 
\ee
where $\mathbf{\Gamma}_{\rm rad}$ is the torque and $\lambda$ is the 
wavelength of the radiation.  

Farsund \& Felderhof (1996) derived expressions for the force and torque
when the incident and scattered waves are expanded in vector spherical 
wave functions.  Since their conventions differ somewhat from ours, we 
briefly review some relevant results.  First, the incident plane wave is
given by
\be
\mathbf{E}_{\rm inc} = \sum_{n, m} \left[ p_{nm} \mathbf{N}_{nm} + q_{nm}
\mathbf{M}_{nm} \right]
\ee
For the TM mode, 
\be
\label{eq:pq_TM}
p_{nm} = - d_{nm} \sin \theta_0 P^{\prime}(\cos \theta_0) \ \ \ \ ; \ \ \ \ 
q_{nm} = d_{nm} m P(\cos \theta_0) / \sin \theta_0
\ee
and for the TE mode, 
\be
p_{nm} = -i m d_{nm} P(\cos \theta_0) / \sin \theta_0 \ \ \ \ ; \ \ \ \
q_{nm} = i d_{nm} \sin \theta_0 P^{\prime}(\cos \theta_0)~~~;
\ee
\be
d_{nm} = i^{n+1} (-1)^m \frac{2n+1}{n (n+1)} \frac{(n-m)!}{(n+m)!}~~~.
\ee

For the TM mode, the coefficients that appear in eqs.~(7.25) 
through (7.30) in Farsund and Felderhof are given by
\be
c_{hnm}^i = \frac{-i m d_{nm} P(\cos \theta_0)}{k_0 \sin \theta_0} 
\ee
\be
c_{enm}^i = \frac{- d_{nm} \sin \theta_0 P^{\prime}(\cos \theta_0)}{k_0}
\ee
\be
c_{hnm}^s = \frac{i \alpha_m a_{n, |m|}(2)}{k_0} \left[ \frac{4\pi}{2n+1}
\frac{(n+|m|)!}{(n-|m|)!} \right]^{1/2}
\ee
\be
c_{enm}^s = \frac{\beta_m a_{n, |m|}(1)}{k_0} \left[ \frac{4\pi}{2n+1}
\frac{(n+|m|)!}{(n-|m|)!} \right]^{1/2}
\ee
with 
$\alpha_m = (-1)^{|m|}$ (-1) for $m < 0$ ($m \ge 0$) and
$\beta_m = (-1)^{|m|}$ (1) for $m < 0$ ($m \ge 0$).

For the TE mode, 
\be
\label{eq:c_TE}
c_{hnm}^i({\rm TE}) = - c_{enm}^i({\rm TM}) \ \ \ \  ;  \ \ \ \ 
c_{enm}^i({\rm TE}) = c_{hnm}^i({\rm TM})
\ee
and the expressions for $c_{hnm}^s$ and $c_{enm}^s$ are the same as for 
the TM mode, except for an additional negative sign when $m<0$. 

To compute the efficiency factors $\mathbf{Q}_{\rm pr}$ and 
$\mathbf{Q}_{\Gamma}$, the expressions in equations (\ref{eq:pq_TM}) through 
(\ref{eq:c_TE}) are substituted into the Farsund \& Felderhof (1996)
expressions for the force and torque (their eqs.~7.25 through 7.30).

\subsection{\label{sec:Q_results} Results}

In Figures \ref{fig:qabs_prolate} through \ref{fig:qtorque_oblate}, 
we display computational results for the efficiency factors, for a prolate
grain, with $\zeta = 3/2$, and an oblate grain, with $\zeta = 2/3$.  We
take $a_{\rm eff} = 0.2 \micron$ and several values of $\theta_0$, the
angle between the incident radiation field and the grain's symmetry axis.
We adopt dielectric functions for ``astronomical silicate'' from 
Draine (2003).  Here, and throughout the paper, we assume unpolarized 
incident radiation; thus, we average over the TE and TM modes.
Although the force has a component perpendicular to $\mathbf{k}_0$, we
only display 
$Q_{\rm pr} \equiv \mathbf{Q}_{\rm pr} \cdot \hat{k}_0$.  
The torque always lies along $\hat{y}$, and vanishes when $\cos \theta_0 = 0$
or 1.  (Recall the definition of the coordinate system, in the paragraph 
following eq.~\ref{eq:hint}.)

For various values of $\zeta$, $\lambda$, and $\theta_0$, we have also 
computed the efficiency factors using the discrete dipole approximation
code DDSCAT (version 6.1, Draine \& Flatau 2004).  In all cases, the results
were identical, to within uncertainties associated with incomplete 
convergence.  We encountered no obstacles in convergence with the PMM 
code, though in some cases we had to take $n_{\rm max} = 80$ to converge
to within $1\%$.  
For some wavelengths, we were unable to converge to within 
$10\%$ with DDSCAT, due to limitations on the size of the dipole arrays
when using the GNU Fortran compiler.  This successful comparison, using
two completely different methods, validates both our PMM code and DDSCAT.

For a given incident radiation field, the spectrally averaged force and torque 
efficiency factors are given by
\be
\mathbf{\bar{Q}}_{\rm pr} = \frac{1}{u_{\rm rad}} \int \mathbf{Q}_{\rm pr} \, 
u_{\nu} \, d\nu
\ee
\be
\label{eq:barQ_Gamma}
\mathbf{\bar{Q}}_{\Gamma} = \frac{1}{\bar{\lambda} \, u_{\rm rad}} 
\int \mathbf{Q}_{\Gamma} \, \lambda \, u_{\nu} \, d\nu~~~,
\ee
where $\nu$ is the frequency, $u_{\nu}$ is the specific energy density, and 
$\bar{\lambda} = \int \lambda \, u_{\nu} d\nu / u_{\rm rad}$. 
In Figure \ref{fig:torque_mmp}, we display $\bar{Q}_{\Gamma, y}$ for
the average interstellar radiation field (ISRF)
spectrum in the solar neighborhood, 
as estimated by Mezger, Mathis, \& Panagia (1982) and Mathis, Mezger, \& 
Panagia (1983).  (See eq.~31 in Weingartner \& Draine 2001b for a convenient 
representation of the ISRF.)  Interestingly, the torque due to radiation
with $\lambda > 1 \micron$ substantially cancels the torque
due to radiation with $\lambda < 1 \micron$.  
For the ISRF, $u_{\rm rad} = 8.63 \times 10^{-13} \erg \cm^{-3}$ and
$\bar{\lambda} = 1.20 \micron$.  

Both the wavelength-dependent and spectrally-averaged efficiency factors
are comparable in magnitude to those for the irregular grain studied by
Draine \& Weingartner (1996); compare figs.~\ref{fig:qtorque_prolate} and
\ref{fig:qtorque_oblate} with their fig.~7 and fig.~\ref{fig:torque_mmp}
with their fig.~12.

For a spherical 
silicate grain with $a_{\rm eff} = 0.2 \micron$ exposed to the ISRF, 
$\bar{Q}_{pr} = 0.82$.  For spheroids with $\zeta = 2/3$ (3/2), 
$\bar{Q}_{\rm pr}$ ranges from 0.78 to 0.89 (0.73 to 0.85), with larger 
values corresponding to larger cross-sectional areas.  The transverse 
component of $\mathbf{Q}_{\rm pr}$ takes values as high as $\approx 0.1$.

\section{Photoelectric Forces and Torques \label{sec:pe}}

\subsection{A Simple Model \label{sec:Fpe_model}}

For simplicity, we will assume that (1) the grains are electrically neutral, 
(2) the photoelectrons emerge along the surface normal $\hat{n}$, and 
(3) the photoelectron energy $E_e = h\nu - W$, where $h\nu$ is the incident
photon energy and $W$ is the work function of the grain material.  
Weingartner \& Draine (2001b) estimated that silicate grains with 
$a_{\rm eff} = 0.2 \micron$ in the cold neutral medium have a slight 
positive charge.  
The rate at which photoelectrons are ejected and the average energy per
photoelectron are both lower for positively charged grains than for neutral
grains.
Thus, each of these simplifications results in an 
overestimate of the force and torque.\footnote{In the rare environments where 
the grains are highly charged, these may be overestimated by over an order of 
magnitude (Weingartner \& Draine 2001a).}
This is appropriate for our 
exploratory study, since we are only attempting to ascertain whether or not
the photoelectric torque may be important enough, compared with the direct
radiative torque, to warrant further consideration in studies of grain
alignment.  We will adopt $W = 8 \eV$ and 
the photoelectric yield $Y$ (i.e., the probability that a photoelectron is
ejected following the absorption of a photon) 
\be
Y = 0.5 \frac{h\nu - W}{5 h\nu - 4 W}~~~;
\ee
these are the estimates of Weingartner \& Draine (2001b) for bulk silicate.
On theoretical grounds, it is expected that the yield and ionization potential
for a grain with $a_{\rm eff} = 0.2 \micron$ deviate little from their 
bulk values (e.g., Watson 1972, 1973).  
However, scant experimental evidence on photoelectric emission
from sub-micron grains is available.  The recent experiment of
Abbas et al.~(2006) found that the yields of such grains substantially 
exceed the bulk yield.     
The work function and yield have not been well characterized experimentally,
even for bulk silicate.
Thus, our results for the force and torque will only be rough estimates, and
we are justified in making the above simplifications.  

Following Kerker \& Wang (1982), we will take the photoemission rate as 
a function of position on the surface proportional to the internal field 
intensity evaluated at the grain surface, 
$|\mathbf{E}_{\rm int}^{\rm surf}|^2$.  With this assumption, as well as
those in the preceding paragraph, 
the photoelectric force $\mathbf{F}_{\rm pe}$ 
and torque $\mathbf{\Gamma}_{\rm pe}$ are given by
\be
\mathbf{F}_{\rm pe} = \pi a_{\rm eff}^2 u_{\rm rad} \mathbf{Q}_{\rm pr, \, pe}
\ee
\be
\mathbf{\Gamma}_{\rm pe} = \pi a_{\rm eff}^2 u_{\rm rad} \frac{\lambda}{2\pi}
\mathbf{Q}_{\Gamma, \, {\rm pe}}
\ee
with
\be
\mathbf{Q}_{\rm pr, \, pe} = \frac{c Q_{\rm abs}}{h\nu} Y p_e 
\mathbf{A}_{\rm pr}
\ee
\be
\mathbf{Q}_{\Gamma, \, {\rm pe}} = \frac{c Q_{\rm abs}}{h\nu} Y p_e 
\frac{2 \pi a_{\rm eff}}{\lambda} \mathbf{A}_{\Gamma}
\ee
\be
\mathbf{A}_{\rm pr} = - 
\frac{\int |\mathbf{E}_{\rm int}^{\rm surf}|^2 \, \hat{n} \, dS}
{\int |\mathbf{E}_{\rm int}^{\rm surf}|^2 \, dS}
\ee
\be
\label{eq:A_Gamma}
\mathbf{A}_{\Gamma} = - \frac{\int |\mathbf{E}_{\rm int}^{\rm surf}|^2 \, 
\mathbf{r} \times \hat{n} 
\, dS} {a_{\rm eff} \int |\mathbf{E}_{\rm int}^{\rm surf}|^2 \, dS}~~~;
\ee
$p_e$ is the photoelectron momentum.

The unit surface normal 
\be
\hat{n} = [1 + (\zeta^4 - 1) \sin^2 \theta]^{-1/2} \, [ \zeta^2 \sin \theta
\, (\hat{x} \cos \phi + \hat{y} \sin \phi) + \hat{z} \cos \theta]
\ee
and the area element
\be
dS = r^2 \, \sin \theta \, \left[ 1 + \left( \frac{r}{a} \right)^4 \, 
\left(1 - \zeta^2 \right)^2 \sin^2 \theta \, \cos^2 \theta \right]^{1/2} \, 
d\theta \, d\phi~~~.
\ee

\subsection{Results}

The photoelectric torque, like the direct radiative torque, always lies along
$\hat{y}$ and vanishes when $\cos \theta_0 = 0$ and 1.  
Figure \ref{fig:a_gamma} displays the $y$-component of the torque asymmetry
factor $\mathbf{A}_{\Gamma}$ for two values of the incident radiation 
wavelength $\lambda$, 
both lying within the relevant $8 \eV$ to $13.6 \eV$ spectral 
range.  

In estimating the maximum value of the photoelectric torque, we
will take $|A_{\Gamma, y}| \approx 0.05$, independent of $\lambda$
(fig.~\ref{fig:a_gamma}).
We will also take $Q_{\rm abs} \approx 1$, a reasonable approximation
when the photon energy exceeds $8 \eV$ (figs.~\ref{fig:qabs_prolate}
and \ref{fig:qabs_oblate}).  Defining the spectrally-averaged efficiency
factor as in equation (\ref{eq:barQ_Gamma}), we find that 
$\bar{Q}_{\Gamma, \, {\rm pe}, \, y} \approx 3.8$.  The energy density in
the ISRF between $8 \eV$ and $13.6 \eV$ is $u(> 8 \eV) = 3.86 \times
10^{-14} \erg \s^{-1}$ and $\bar{\lambda} = 0.124 \micron$ for this
spectral range.  From Figure \ref{fig:torque_mmp}, 
$|\bar{Q}_{\Gamma, y}|$ reaches values as high as $\approx 0.05$ for 
the ISRF.  Comparing the radiative and photoelectric torques at the 
$\theta_0$ for which they peak, we find that 
$\Gamma_{\rm pe} \approx 0.35 \, \Gamma_{\rm rad}$.  

If the photoelectric yield varies across the grain surface, then the torque
is no longer confined to lie along $\hat{y}$, raising the possibility 
that even symmetric shapes like spheroids could experience radiation-driven
alignment (Purcell 1979).  In this case, equation 
(\ref{eq:A_Gamma}) for $\mathbf{A}_{\Gamma}$ is modified, with the yield $Y$ 
appearing in both integrals.  The $z$-component of $\mathbf{A}_{\Gamma}$
remains zero, since the $z$-component of
$\mathbf{r} \times \hat{n}$ always vanishes.  Specifically, we situated a
single large spot on the grain surface where the yield is enhanced by 
$10\%$.  When the spot is on the illuminated side of the grain, 
$A_{\Gamma, \, y}$ changes by as much as $\approx 20\%$  and 
$|A_{\Gamma, \, x}|$  (and $|A_{\Gamma, \, y}|$ for $\cos \theta_0 = 0$, 1)
reach values as high as $\approx 0.005$.  Thus, for real interstellar grains,
the aligning torque associated with non-symmetric shape likely dominates
any aligning torque associated with non-uniform photoelectric yield.

The forward-direction force asymmetry parameter 
$A_{\rm pr} \equiv \mathbf{A}_{\rm pr} \cdot \hat{k}_0 \approx 0.4$ 
to 0.6, while that for the transverse direction ranges from zero to 
$\approx 0.15$.  With $A_{\rm pr} \approx 0.5$ and $Q_{\rm abs} \approx 1$,
the ISRF spectrally-averaged efficiency factor
$\bar{Q}_{\rm pr, \, pe} \approx 3.7$, yielding 
$F_{\rm pe} \approx 0.2 \, F_{\rm rad}$.

\section{Photodesorption \label{sec:photodesorption}}

Although uncertainties abound in modelling the photoelectric force 
and torque, photodesorption presents even greater challenges, as described
in \S 4 of Weingartner \& Draine (2001a).  A major question is whether
or not adsorbed H atoms can diffuse across the surface, by either thermal
barrier hopping or quantum mechanical tunneling.  This depends on the 
poorly known surface-adatom binding energy.  

For uniform surface properties, the magnitude of the photodesorption torque
increases with increasing surface coverage of adatoms.  Since we seek to 
estimate the maximum plausible photodesorption torque, we assume complete
coverage across the entire surface.  This results, for example, if the 
adatoms do not desorb and the rate at which gas-phase atoms collide with 
the grain greatly exceeds the removal rates (due to H$_2$ formation as
well as photodesorption).  

Photodesorbed H atoms are produced at a rate $\ltsim R_{\rm pd}^0 S / l^2$, 
where $R_{\rm pd}^0$ is the photodesorption rate per adatom, $S$ is the grain 
surface area, and $l^2$ is the surface area per binding site.  We adopt the 
crude estimate from Weingartner \& Draine (2001a) of $R_{\rm pd}^0 \approx
2 \times 10^{-10} \s^{-1}$ for grains exposed to the ISRF, and we take
$l^2 \approx 10 \Angstrom^2$.  We also assume that the kinetic energy of
photodesorbed atoms is comparable to that of photoelectrons, i.e., 
$\sim \eV$.  Finally, we assume that the photodesorption rate as a function
of position on the surface is proportional to the electric intensity just 
above the surface.

The ratio of the photodesorption torque $\mathbf{F}_{\rm pd}$ to the
photoelectric torque can then be simply estimated as
\be
\frac{F_{\rm pd}}{F_{\rm pe}} \sim \frac{S}{\pi a_{\rm aeff}^2} 
\frac{R_{\rm pd}^0 h \nu}{c u_{\rm rad} Q_{\rm abs} Y l^2} \left(
\frac{m_{\rm H}}{m_e} \right)^{1/2} \frac{|\mathbf{A}_{\Gamma}({\rm pd})|}
{|\mathbf{A}_{\Gamma}({\rm pe})|}~~~,
\ee
where $S \approx 4 \pi a_{\rm eff}^2$ for $\zeta = 3/2$ and 2/3, 
$m_{\rm H}$ is the proton mass, $\mathbf{A}_{\Gamma}({\rm pe})$ is
the asymmetry parameter given in equation (\ref{eq:A_Gamma}), and 
$\mathbf{A}_{\Gamma}({\rm pd})$ is identical, except that the internal 
electric field $\mathbf{E}_{\rm int}^{\rm surf}$ is replaced with 
the external electric field at the surface, 
$\mathbf{E}_{\rm ext}^{\rm surf}$.  For $\lambda = 0.15 \micron$, 
$A_{\Gamma, y}({\rm pd})$ differs very little from 
$A_{\Gamma, y}({\rm pe})$, but for $\lambda = 0.1 \micron$, the external
asymmetry factor is as much as three times smaller than the internal
asymmetry factor.  With $Q_{\rm abs} \approx 1$, $Y \approx 0.07$, and
$A_{\Gamma}({\rm pd}) \approx A_{\Gamma}({\rm pe})$, 
$\Gamma_{\rm pd} \approx 0.3 \, \Gamma_{\rm pe}$.  

The external and internal pressure asymmetry factors are also generally 
comparable, though the former can be as much as a factor of 2 smaller than
the latter.  Thus, for the models considered here, $F_{\rm pd}$ and 
$F_{\rm pe}$ are of comparable magnitude, as Weingartner \& Draine (2001a)
found for spheres.

\section{Conclusion \label{sec:conclusion}}

The goal of this work was to determine whether or not recoil torques
must be included in investigations of radiation-driven grain alignment.
Our model for the photoelectric torque neglected a few effects that would
tend to suppress the torque (\S \ref{sec:Fpe_model}).  We found 
$\Gamma_{\rm pe} \approx 0.35 \, \Gamma_{\rm rad}$, suggesting that 
the photoelectric torque may actually contribute at about the 10\% level 
compared with the direct radiative torque.

However, the photoelectric yield for sub-micron silicate grains is 
highly uncertain.  If the recent experimental results of Abbas et al.~(2006)
are accurate, then the yield may be substantially larger than we have 
assumed.  This would increase the torque, but not in direct proportion, 
since higher yield implies higher positive grain charge, which partially 
suppresses the torque.  Yet, we cannot rule out the possibility that the 
photoelectric torque is of the same order of magnitude as the radiative
torque for spheroids.
Clearly, additional experiments are needed to better characterize 
photoelectric emission from sub-micron silicate grains.

\begin{figure}
\epsscale{1.00}
\plotone{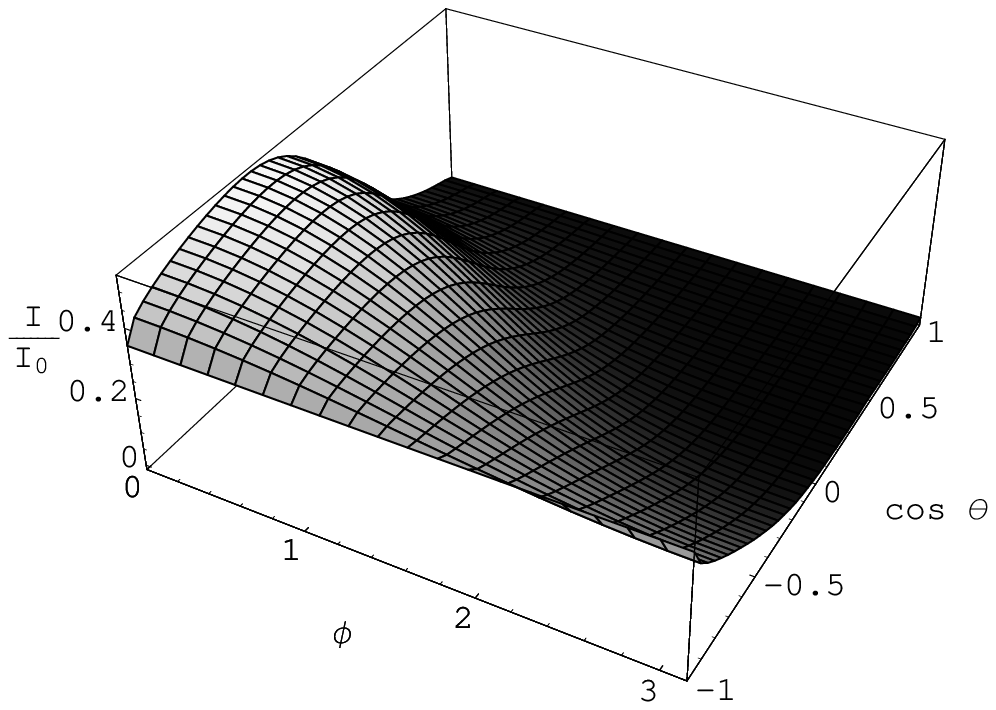}
\caption{
\label{fig:oblate.0.1.int.4}
The intensity $I = |\mathbf{E}|^2$ of the internal field (averaged over the TE 
and TM modes
and normalized to the incident intensity $I_0$) at the grain surface, for an 
oblate ($\zeta = 2/3$) silicate grain with $a_{\rm eff} = 0.2 \micron$, 
incident wavelength $\lambda = 0.1 \micron$, and angle between the incident
radiation and the grain symmetry axis $\theta_0 = \cos^{-1} 0.6$.  
        }
\end{figure}

\begin{figure}
\epsscale{1.00}
\plotone{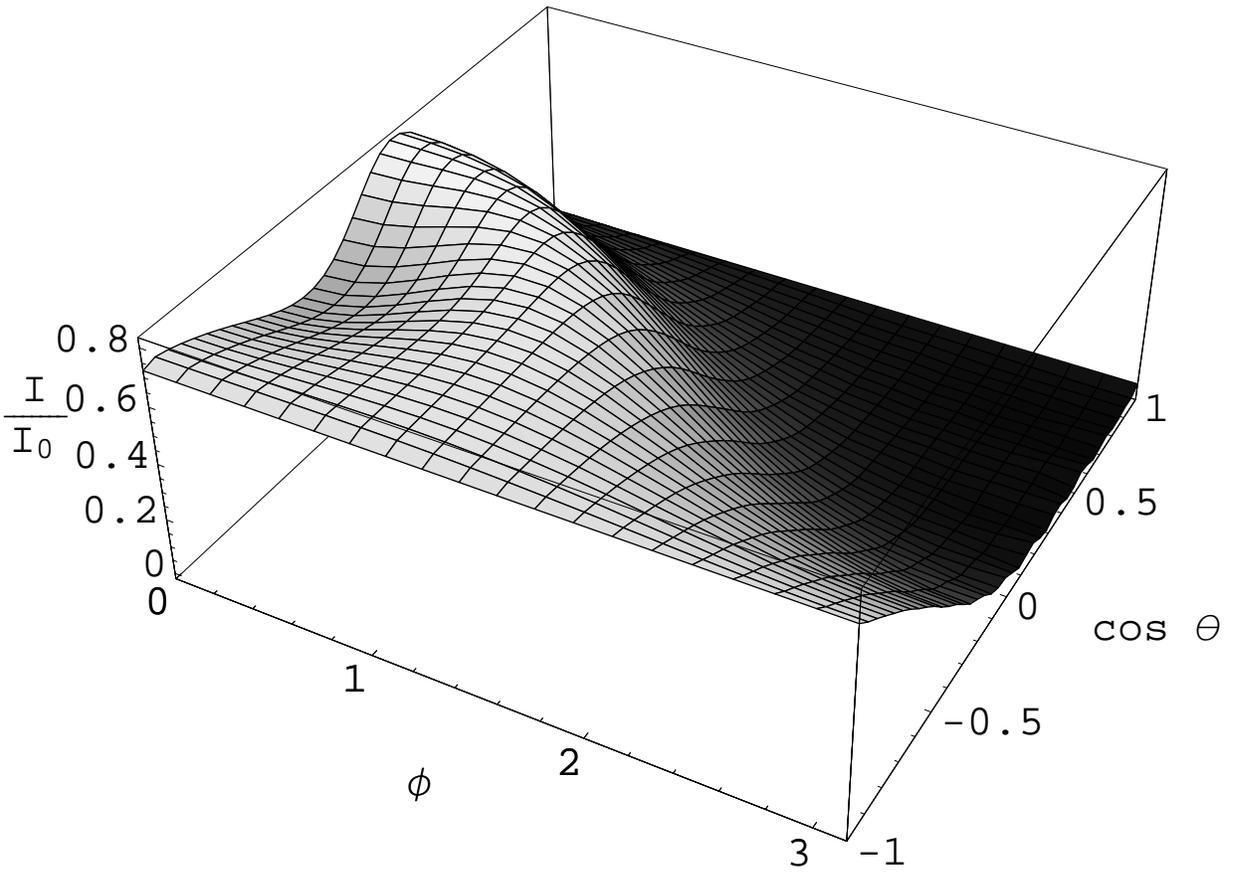}
\caption{
\label{fig:oblate.0.1.ext.4}
Same as fig.~\ref{fig:oblate.0.1.int.4}, except that the external field
intensity at the grain surface is plotted.
        }
\end{figure}

\begin{figure}
\epsscale{1.00}
\plotone{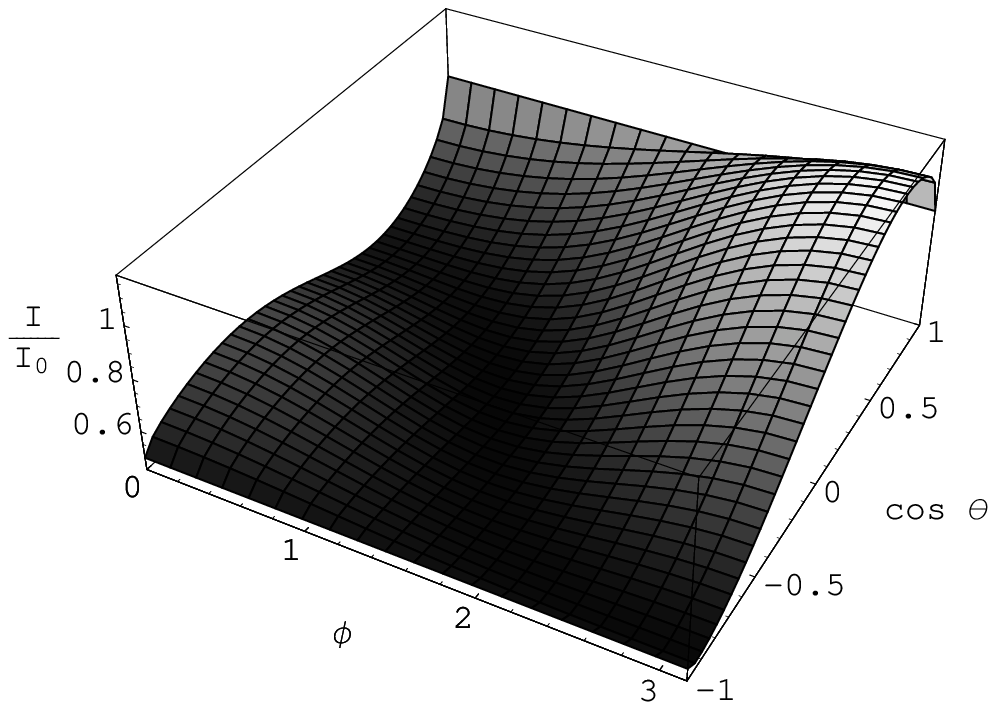}
\caption{
\label{fig:oblate.1.0.int.4}
Same as fig.~\ref{fig:oblate.0.1.int.4}, except that $\lambda = 1 \micron$.
        }
\end{figure}

\begin{figure}
\epsscale{1.00}
\plotone{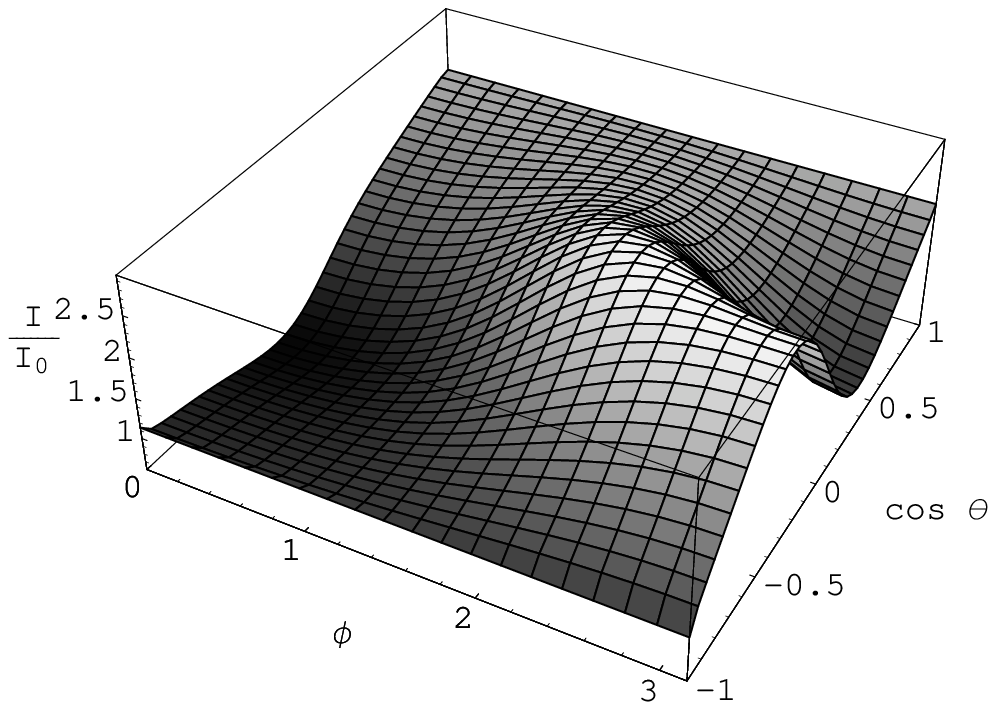}
\caption{
\label{fig:oblate.1.0.ext.4}
Same as fig.~\ref{fig:oblate.0.1.ext.4}, except that $\lambda = 1 \micron$.
        }
\end{figure}

\begin{figure}
\epsscale{1.00}
\plotone{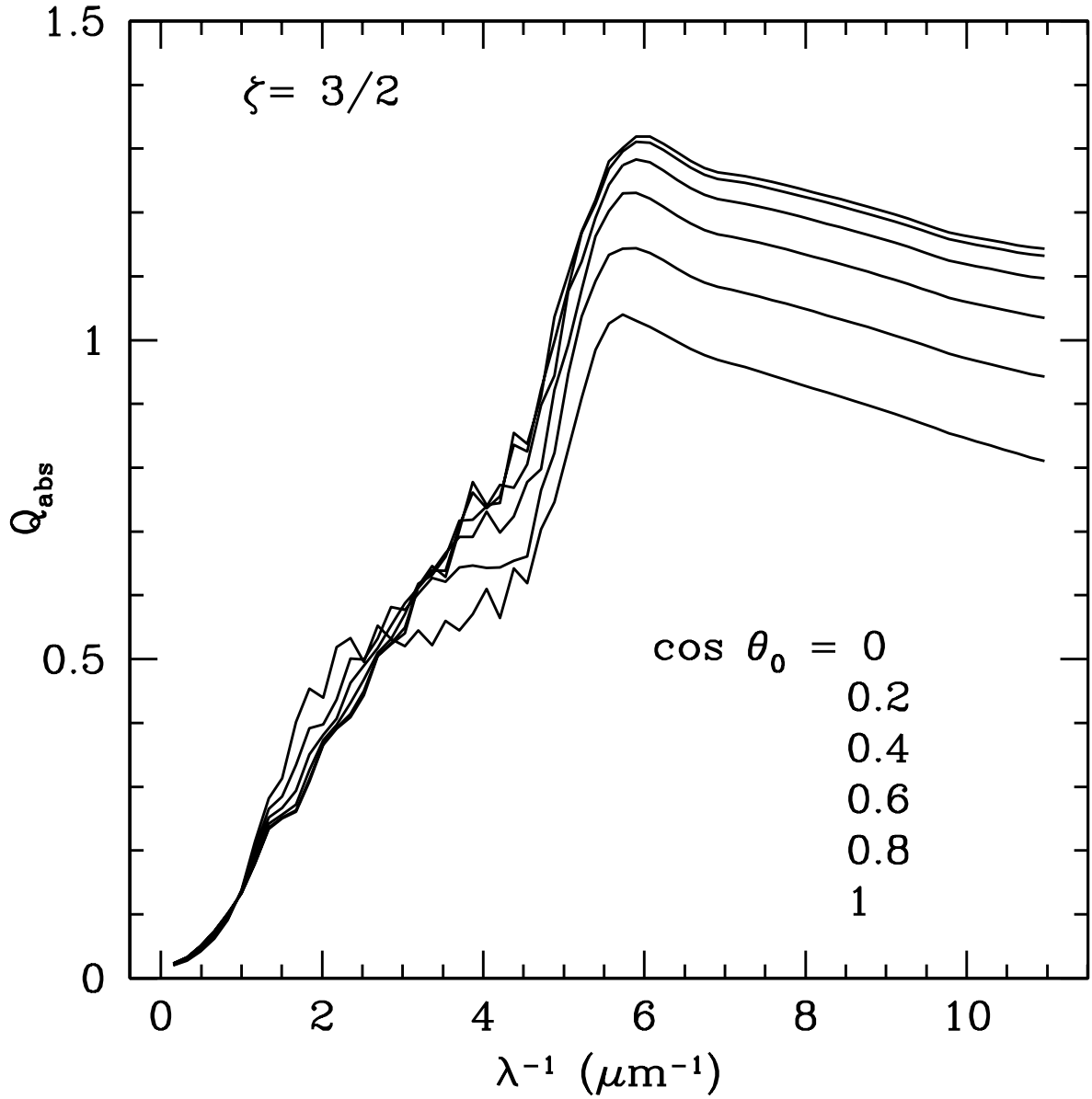}
\caption{
\label{fig:qabs_prolate}
$Q_{\rm abs}$ for prolate ($\zeta = 3/2$) grains with 
$a_{\rm eff} = 0.2 \micron$ and various values of $\theta_0$.  
        }
\end{figure}

\begin{figure}
\epsscale{1.00}
\plotone{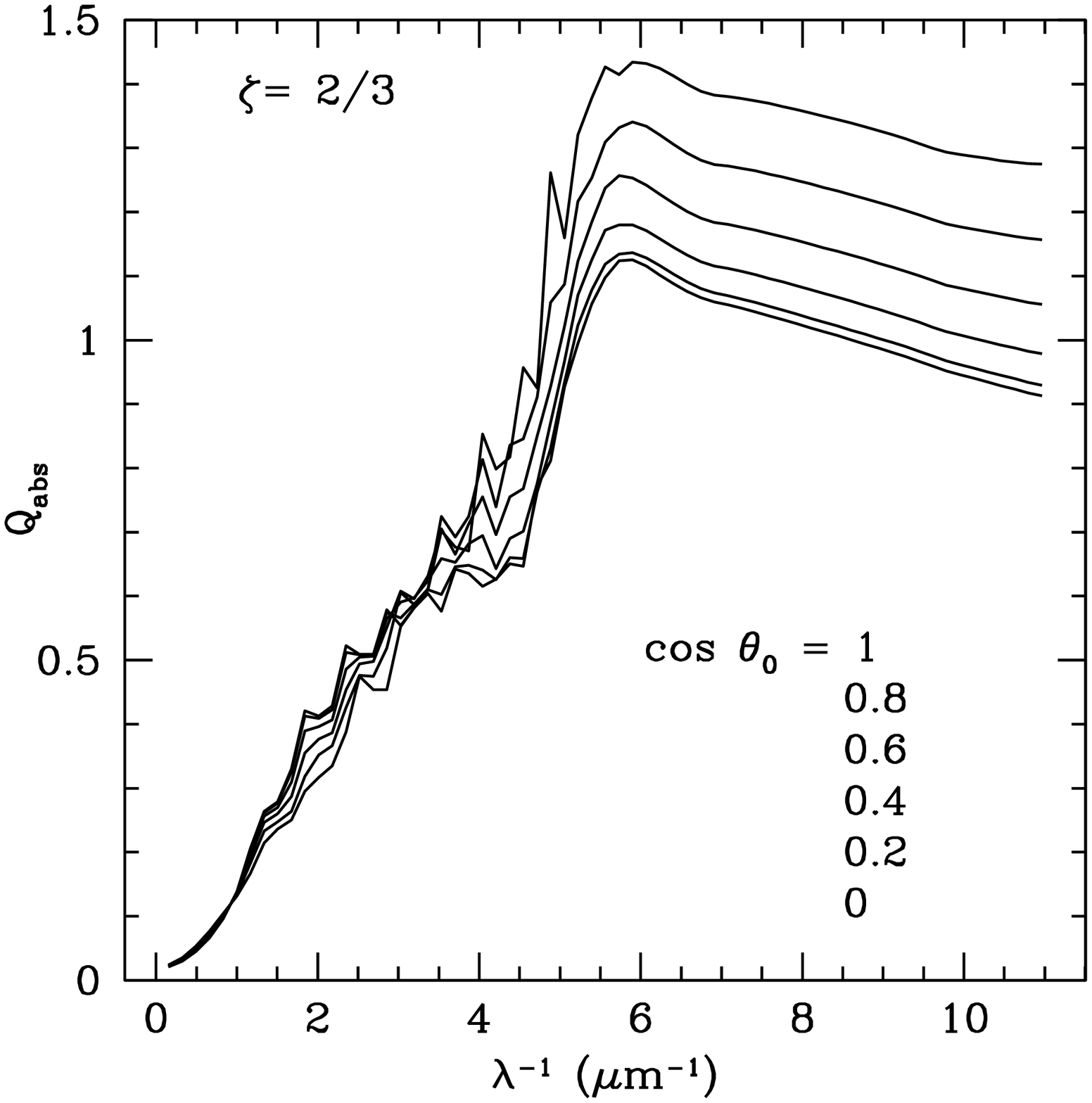}
\caption{
\label{fig:qabs_oblate}
$Q_{\rm abs}$ for oblate ($\zeta = 2/3$) grains 
with $a_{\rm eff} = 0.2 \micron$.
        }
\end{figure}

\begin{figure}
\epsscale{1.00}
\plotone{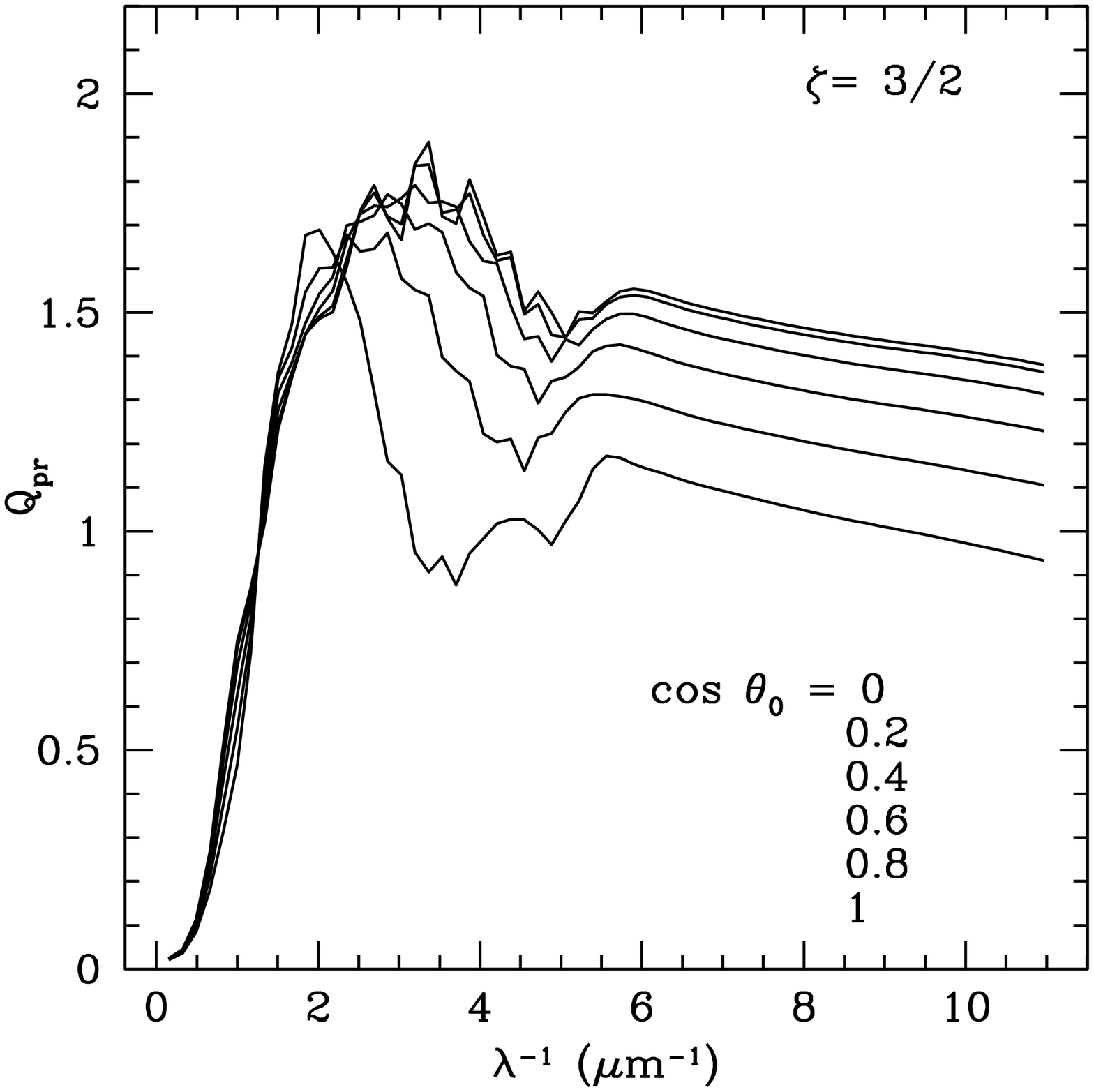}
\caption{
\label{fig:qpr_prolate}
$Q_{\rm pr}$ for prolate ($\zeta = 3/2$) grains
with $a_{\rm eff} = 0.2 \micron$.
        }
\end{figure}

\begin{figure}
\epsscale{1.00}
\plotone{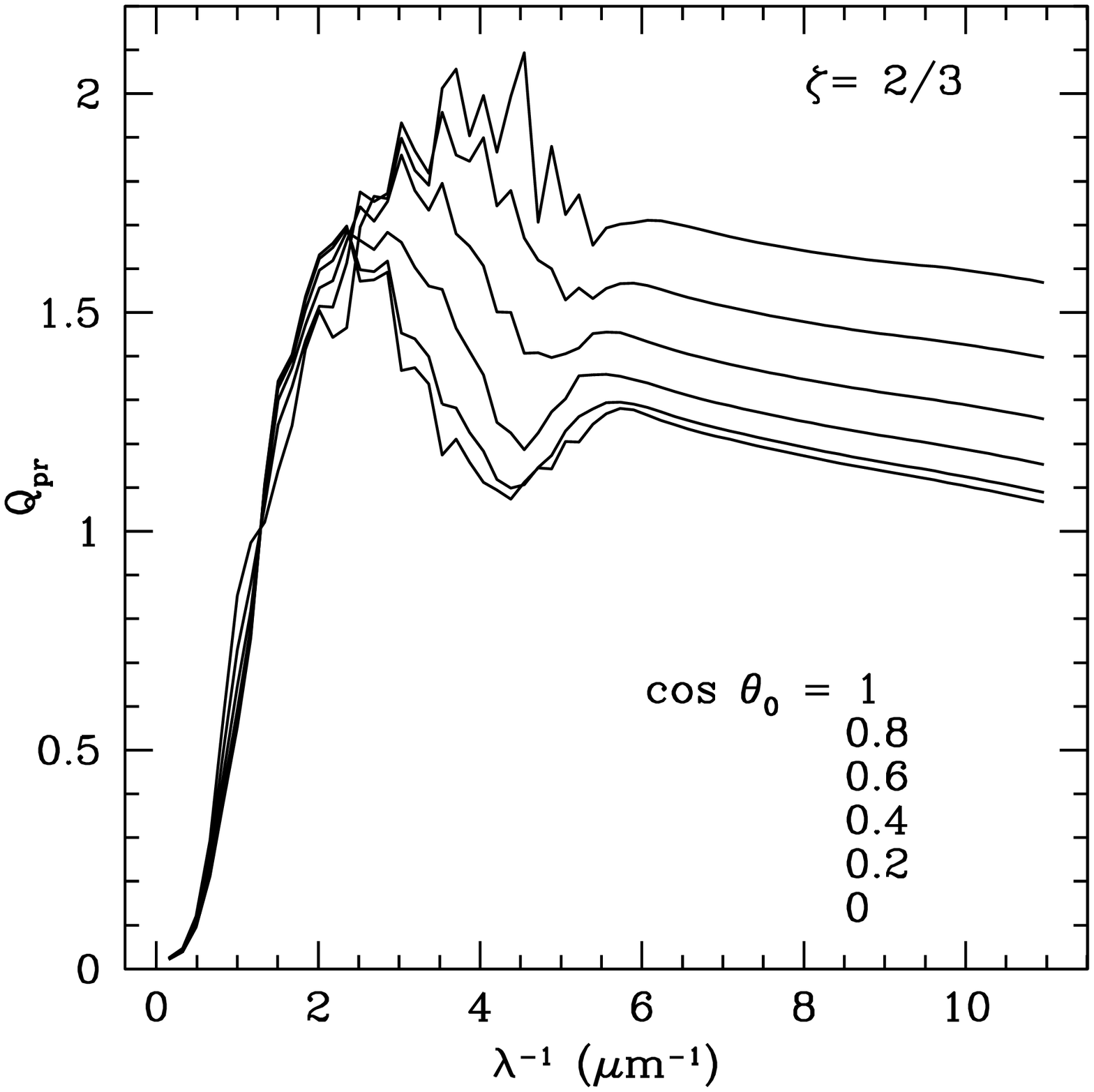}
\caption{
\label{fig:qpr_oblate}
$Q_{\rm pr}$ for oblate ($\zeta = 2/3$) grains
with $a_{\rm eff} = 0.2 \micron$.
        }
\end{figure}

\begin{figure}
\epsscale{1.00}
\plotone{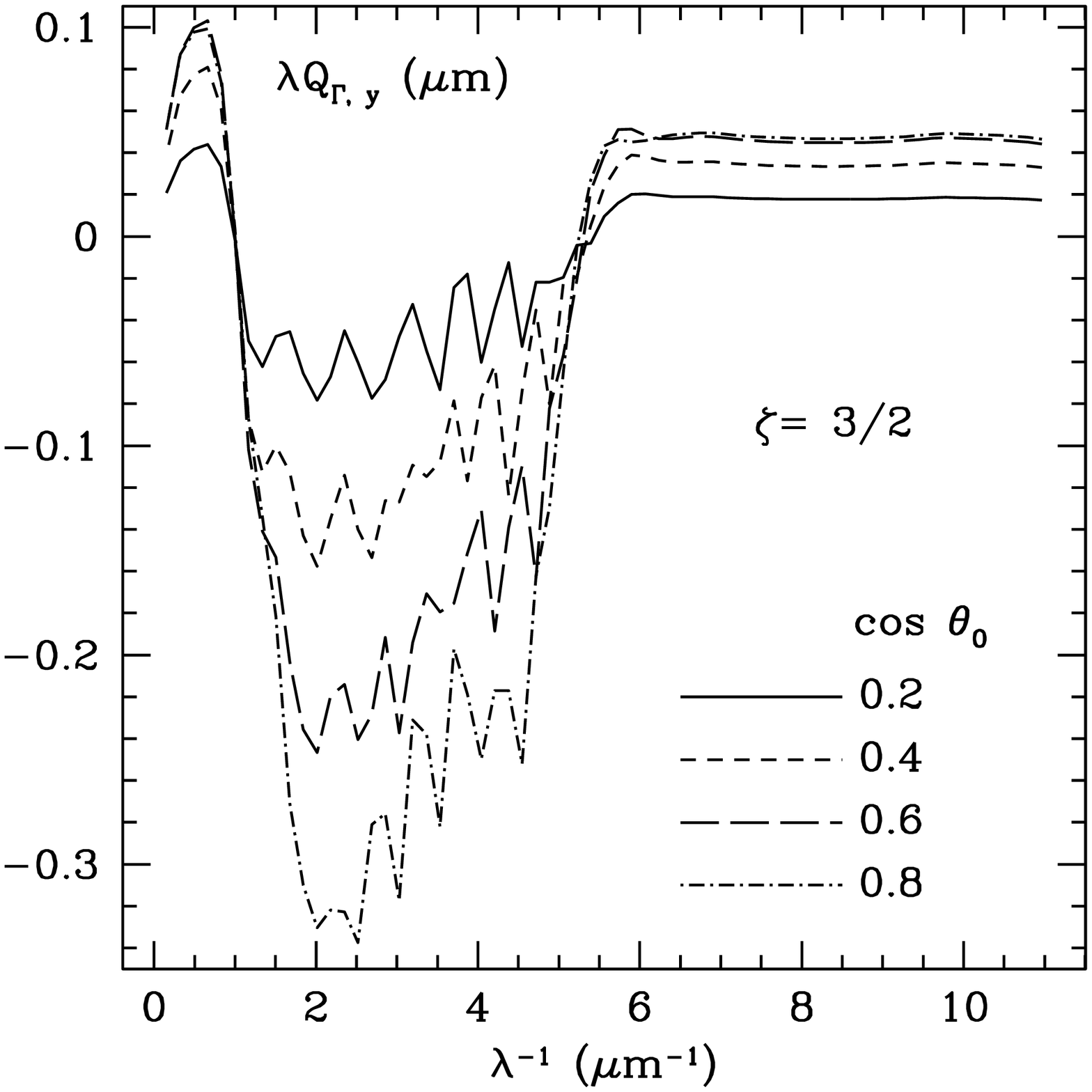}
\caption{
\label{fig:qtorque_prolate}
$\lambda Q_{\Gamma, y}$ for prolate ($\zeta = 3/2$) grains
with $a_{\rm eff} = 0.2 \micron$.
        }
\end{figure}

\begin{figure}
\epsscale{1.00}
\plotone{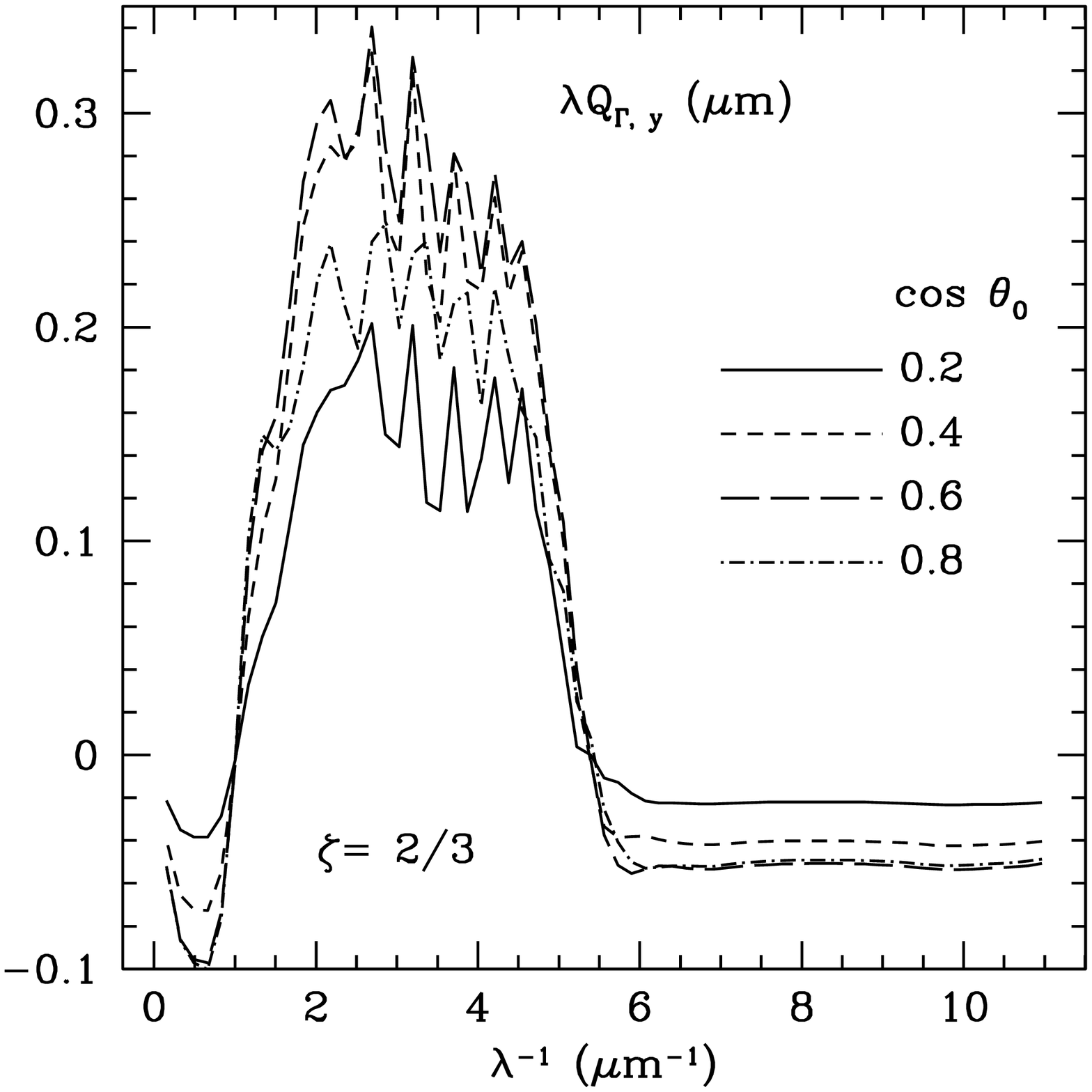}
\caption{
\label{fig:qtorque_oblate}
$\lambda Q_{\Gamma, y}$ for oblate ($\zeta = 2/3$) grains
with $a_{\rm eff} = 0.2 \micron$.
        }
\end{figure}

\begin{figure}
\epsscale{1.00}
\plotone{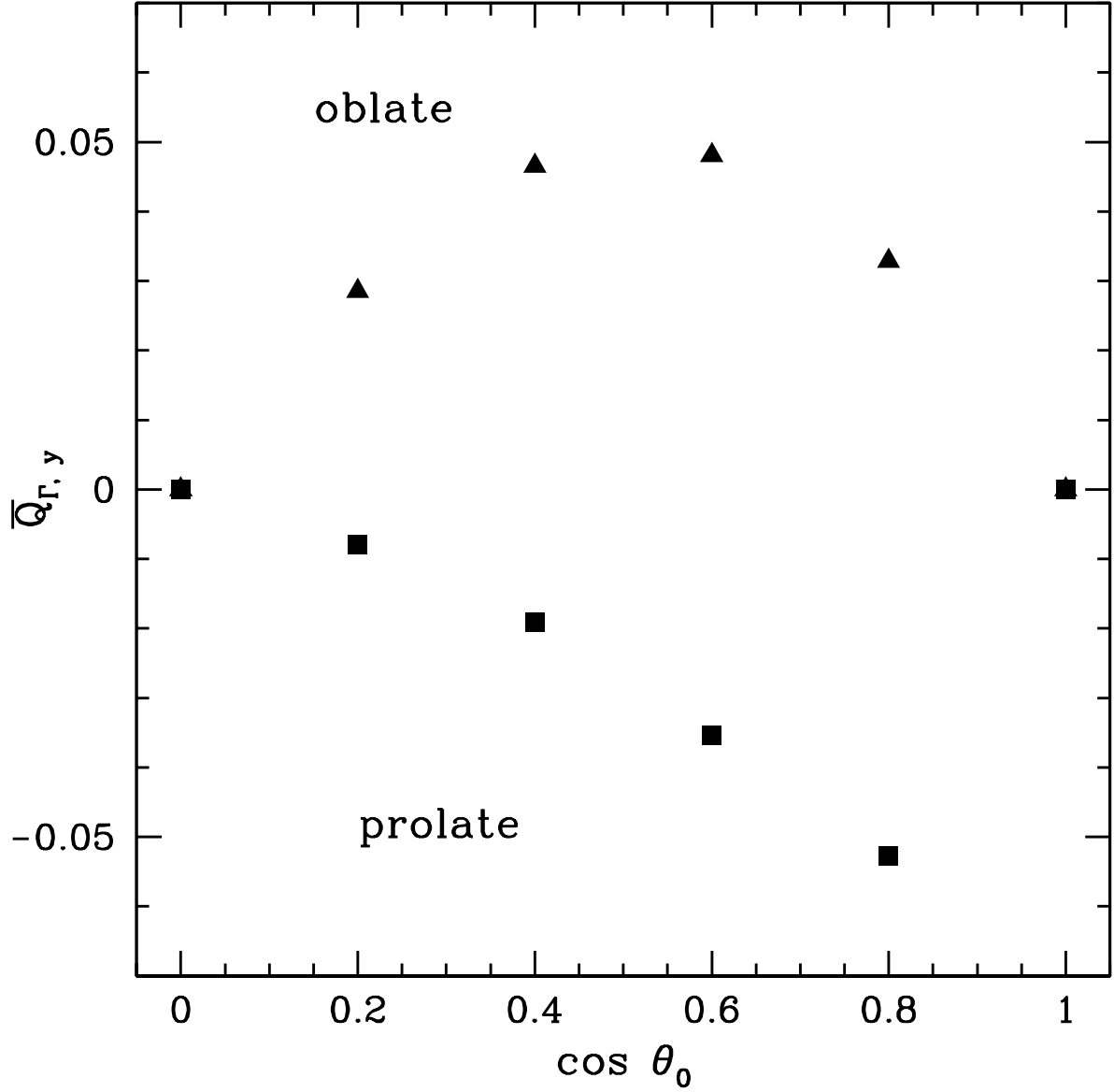}
\caption{
\label{fig:torque_mmp}
The torque efficiency factor averaged over the MMP radiation field, 
$\bar{Q}_{\Gamma, y}$, for prolate ($\zeta = 3/2$) and oblate 
($\zeta = 2/3$) grains with $a_{\rm eff} = 0.2 \micron$.
        }
\end{figure}

\begin{figure}
\epsscale{1.00}
\plotone{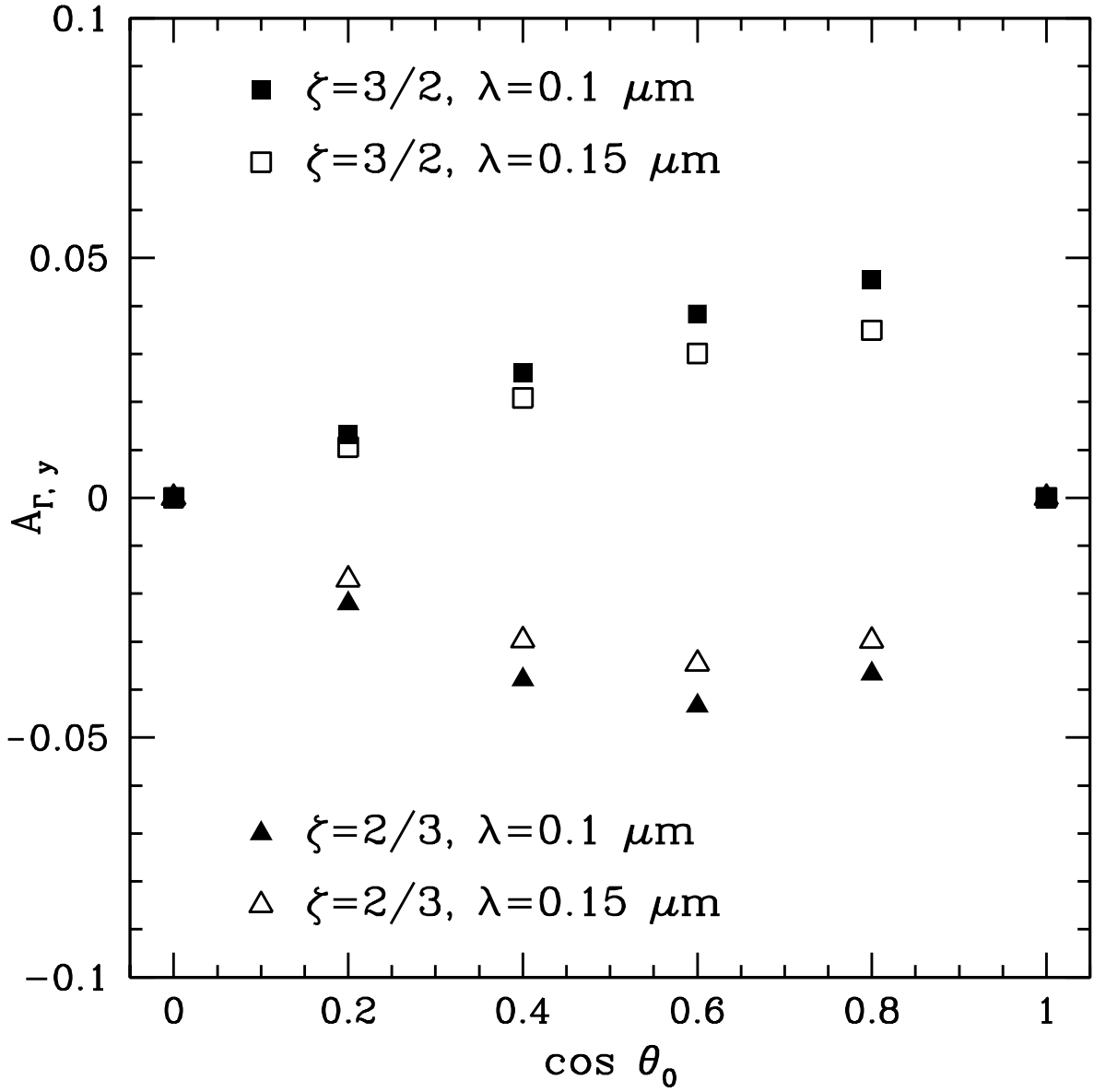}
\caption{
\label{fig:a_gamma}
The $y$-component of the torque asymmetry factor $\mathbf{A}_{\Gamma}$.
        }
\end{figure}

We must also acknowledge the possibility that less symmetric grains may 
experience larger recoil torques than spheroids, even though the direct
radiative torques for the spheroids examined here are comparable in magnitude
to that for the irregular grain studied by Draine \& Weingartner (1996).
The computational demands associated with the PMM are much more severe for
grains lacking azimuthal symmetry.  However, we have found that the 
near-surface fields for spheroids are extremely smooth 
(figs.~\ref{fig:oblate.0.1.int.4} through \ref{fig:oblate.1.0.ext.4}),  
suggesting that the 
discrete dipole approximation may yield sufficient resolution for accurate 
evaluation of the recoil torques.  We will pursue this possibility in future 
work.

The photodesorption torque appears to contribute at most at the 10\% level,
although it may be more important if our adopted value for the photodesorption
rate, $R_{\rm pd}^0$, is too small.  In making our simple estimate, we 
assumed that the surface-adatom binding energy is relatively large 
($> 1 \eV$), but smaller binding energies (as suggested by, e.g., the 
recent experiment of Perets et al.~2007) may result in a smaller torque.  
Futher experimental work characterizing the surfaces of amorphous 
silicates, and the photodesorption rates, is needed before the torque can 
be reliably estimated.  Thus, for now, we recommend that both recoil 
torques be omitted from detailed studies of radiation-driven alignment.

\acknowledgements

We are grateful to Bruce Draine and an anonymous referee for helpful 
comments.  
JCW is a Cottrell Scholar of Research Corporation.
Support for this work, part of the Spitzer Space Telescope Theoretical
Research Program, was provided by NASA through a contract issued by the
Jet Propulsion Laboratory, California Institute of Technology under a
contract with NASA.

\end{document}